\documentclass[%
reprint,
superscriptaddress,
amsmath,amssymb,
aps,
prf,
floatfix
]{revtex4-2}

\usepackage{graphicx}
\usepackage{dcolumn}
\usepackage{bm}
\usepackage{comment}%
\usepackage{amsmath}%
\usepackage{url}%
\usepackage[mathlines,pagewise]{lineno}

\begin{document}
	
	\preprint{APS/123-QED}
	
	\title{Unifying Lengthscale-Based Rheology of Dense Granular-Fluid Mixtures}
	\author{Zhuan Ge}
    \affiliation{College of Civil Engineering and Architecture, Zhejiang University, 866 Yuhangtang Road, Hangzhou 310058, Zhejiang, China\\}
\affiliation{ Key Laboratory of Coastal Environment and Resources of Zhejiang Province (KLaCER), School of Engineering, Westlake University, 18 Shilongshan Street, Hangzhou, Zhejiang 310024, China.\\}

	\author{Teng Man}%
	\email{manteng@westlake.edu.cn}
\affiliation{ Key Laboratory of Coastal Environment and Resources of Zhejiang Province (KLaCER), School of Engineering, Westlake University, 18 Shilongshan Street, Hangzhou, Zhejiang 310024, China.\\}
\author{Herbert E. Huppert}%
\affiliation{
	Institute of Theoretical Geophysics, King's College, University of Cambridge, King's Parade, Cambridge CB2 1ST, United Kingdom\\
}%

\author{Kimberly Hill}%
\email{kmhill@umn.edu}
\affiliation{
	Department of Civil, Environmental, and Geo-Engineering, University of Minnesota, Minneapolis, Minnesota, USA
}%
\author{Sergio Andres Galindo-Torres}
\email{s.torres@westlake.edu.cn}
\affiliation{ Key Laboratory of Coastal Environment and Resources of Zhejiang Province (KLaCER), School of Engineering, Westlake University, 18 Shilongshan Street, Hangzhou, Zhejiang 310024, China.\\}
\date{\today}
\begin{abstract}
In this communication, we present a new lengthscale-based rheology for dense sheared particle suspensions as they transition from inertial- to viscous-dominated.  We derive a lengthscale ratio using straightforward physics-based considerations for a particle subjected to pressure and drag forces.  In doing so, we demonstrate that an appropriately chosen length-scale ratio intrinsically provides a consistent relationship between normal stress and system proximity to its ''jammed'' or solid-like state, even as a system transitions between inertial and viscous states, captured by a variable Stokes number.   
\end{abstract}

\maketitle


Particle flows and particle-fluid are ubiquitous in natural phenomena, such as landslides, debris flows, and rock falls \cite{1995The,2002Avalanche,2020Pore}.
Complex environmental conditions make it difficult to obtain a unified constitutive law for their flow characteristics, particularly for dense flows, where short-range interactions are enduring and often generate long-range correlations\cite{kamrin2019non}.

In the last two decades, significant progress has been made in modeling the flows of wet and dry granular flows focused on dimensional analysis and time scales. For dry granular flows, the local normal stress $P_{p}$ (which is associated with interparticle interactions only), particle density $\rho_{s}$, particle size $d$, and shear rate $\dot{\gamma}$ are combined into a single dimensionless ratio of two timescales, microscopic ($\sqrt{\rho_{s}d^{2}/P_{p}}$) and macroscopic ($1/\dot{\gamma}$): $I=\dot{\gamma}d/\sqrt{P_{p}/\rho_{s}}$  \cite{GDR2004On,da2005rheophysics,jop2006constitutive}.  Various authors have shown that dynamic parameters such as the apparent friction coefficient $\mu=\tau / P_{p}$ (here, $\tau$ is a local shear stress) and the solid fraction $\phi$ can be represented using functions of $I$ only including steady-state \cite{GDR2004On} and transient (e.g., column collapse) systems \cite{lacaze2009axisymmetric}. Cassar et al.\cite{cassar2005submarine} adapted this framework to saturated particle systems by replacing the microscopic (inertial) timescale with a viscous timescale ($P_p/\eta_f$, where $\eta_{f}$ is the fluid viscosity), and the appropriate dimensionless control parameter is $J=\dot{\gamma} \eta_{f}/P_{p}$. Boyer et al.\cite{boyer2011unifying} further validated the saturated framework and generalized the form to include much sparser suspensions. 

In the last decade, work on dense flows has broadened to include systems in which both fluid viscous forces and particle inertial effects contribute to the rheology. Toward this, Trulsson et al. (2012) \cite{trulsson2012transition} proposed using effective shear stress taking the form of superposed inertial and viscous stresses ($\tau_{\rm{eff}}=\lambda \times \rho_{s} d^2\dot{\gamma}^2+\eta_f \dot{\gamma}$ normalized by $P_p$ yielding a new dimensionless number:  $K=\lambda I^{2}+J$  ($\lambda$ is a single-valued fit parameter).   Tapia et al.\cite{Tapia_PhysRevLett.129.078001} argued that $\lambda=1/St_{\rm{tr}}$, where $St_{\rm{tr}}$ is a transitional Stokes number ($St=I^2/J$) close to 1. Specifically, for saturated 3-d experiments, they found $St_{\rm{tr}}=10$ provided a reasonable collapse for their data. 

These efforts have revolutionized the representation of dense granular-fluid flows, remarkable in their form of dimensional analysis and data fitting.  Nevertheless, a significant issue remains\cite{man2018rheology}, associated partly with the dimensional focus of the work.  As noted by Bagnold (1954)\cite{bagnold1954experiments}, fundamental physical considerations indicate that an additional (dimensionless) variable is needed, for example, to represent the relative importance of inertial and viscous contributions \textit{as they vary} from one system to the next.  

In this communication, we use theoretical considerations to find a more consistent physics-based characterization of the details underlying the transition between inertial and viscous flows. Through these efforts, we find consideration of length scale ratios provides substantial insights to the problem and, at the same time, a framework consistent with that proposed originally by Trulsson, Tapia, and colleagues.  Specifically, in the expression for $K$ above, a function $\lambda_{\rm{St}}$ that increases with $St$ replaces the constant $\lambda$. Beyond this, we note that the lengthscale considerations that lead us to this framework are physically more well-suited to represent the dynamics close to maximum concentration for flow: $\phi_c$ (e.g., near jamming).  In particular, these results suggest that lengthscales are intrinsically related to the limited particulate movements as they approach jamming and thus can capture commonly reported the influence of the proximity of  $\phi/\phi_c$ to $1$ to the influence of particle displacements\cite{bocquet2009kinetic} and other system dynamics.  As evidence, we demonstrate that the new framework captures previously published data\cite{savage1984289,WOS:000331956500014,boyer2011unifying,amarsid2017viscoinertial,Tapia_PhysRevLett.129.078001} and new computational data presented herein.  
We begin by considering representative forces on a particle and use this to derive lengthscale and timescale ratios of the problem.  As noted by Cassar et al., (2005) \cite{cassar2005submarine}, when inertia, drag force, and normal stress via interparticle contacts $P_p$ are all important, we may write:  
\begin{equation}
    (\pi/6)\rho_{s}d^3\frac{d v}{d t}=(\pi/4) P_p d^2-F_{d},
    \label{eq:equli}
\end{equation}
A literal interpretation of this formula involves the response of a particle to an interparticle contact on one side supplying a representative of the local normal stress ($P_p$) and no particle contact on the other side (e.g., due to a ``hole'' in the contact network).  For simplicity, we approximate the drag force by the Stokes force, i.e., proportional to speed $v$: $F_{d}=3\pi\eta_{f} vd$. We consider a particle released from rest, and integrate this expression to find first velocity $v(t)$ and then a relationship between distance travelled and time: 

\begin{figure}
	\centering
	\includegraphics[scale=0.235]{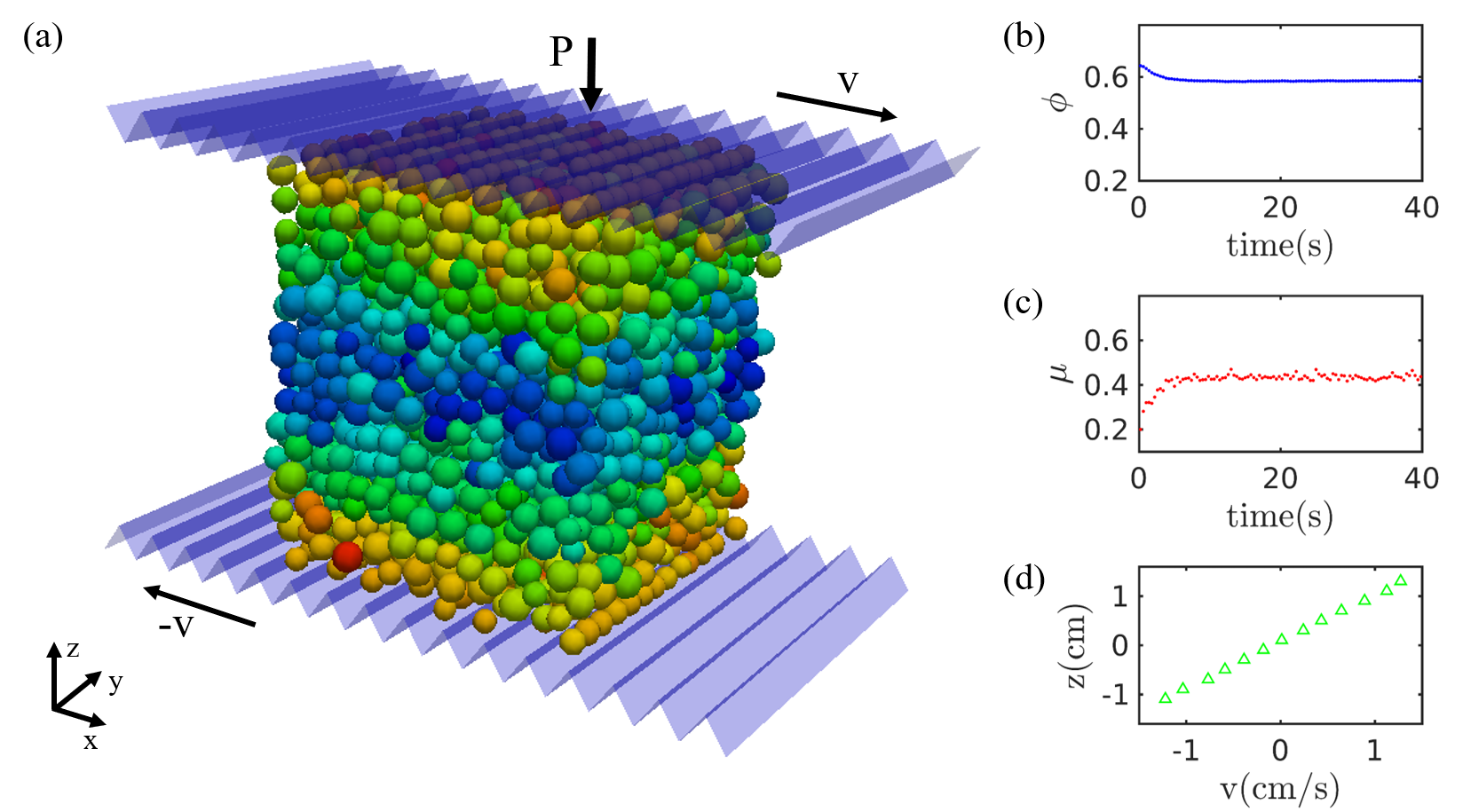}
	\caption{(a) Planar shear simulation setup. The top plate exerts a pressure $P_{p}$ on the particles while the granular assembly is sheared by the top and bottom plates under the same velocity $v$ in the opposite direction. Samples of (b) the solid fraction $\phi$ of the granular assembly, and (c) the apparent friction coefficient $\mu$ of the granular assembly for the simulation under fluid. (d) velocity distribution of particles in the Z direction at time=40s to illustrate the approach of the system to a steady state when $v(z)$ is linear and $\mu$ and $\phi$ are unchanging.  }
	\label{setup}
\end{figure}

\begin{subequations}
\begin{equation}
    v(t)=\int_{t'=0}^{t} [dv(t')/dt']dt' = v_{\rm{f}}(1-e^{-at}),
    \label{eq:v(t)}
\end{equation} 
\begin{equation}
l(t)=\int_{t'=0}^{t} v(t')dt' = v_{\rm{f}}t-\frac{v_{\rm{f}}}{a}(1-e^{-at}).\label{eq:l(t)}
\end{equation}
\end{subequations}
$v_{\rm{f}}=Pd/(12\eta_{\rm{f}})$ is analogous to a settling velocity, and $ a=18\eta_{f}/(\rho_{s}d^2)$, to a settling timescale. We can use Eqn.\ \ref{eq:l(t)} to derive either a  ``microscopic timescale'' $t_{\mu}$ for a particle to travel distance $l=d$ or a ``microscopic lengthscale'' $l_{\mu}$ a particle can travel in time $\mathcal{T}=1/\dot{\gamma}$:  
\begin{subequations}
\begin{equation}
    t_{\mu}=\frac{12\eta_{f}}{P_{p}}+\frac{\rho_{s}d^2}{18\eta_{f}}\left[1+W\left(-e^{-216\eta_{f}^2/(\rho_{s}d^2P_{p})-1}\right)\right]
\end{equation} \label{eq:t(mu)}
\begin{equation}
l_{\mu}=\frac{v_{\rm{f}}}{\dot{\gamma}}-\frac{v_{\rm{f}}}{a}(1-e^{-a/\dot{\gamma}})
\end{equation}
\end{subequations}
where $W(\cdot)$ is the Lambert-W function. (See supplement for more details of this and other calculations.)

With these, we find a dimensionless timescale ratio by dividing $t_{\mu}$ by the (``macroscopic'') timescale  $\mathcal{T}=1/\dot{\gamma}$:  
\begin{equation}
            G=\frac{t_{\mu}}{\mathcal{T}}=12J+\frac{\rm{St}}{18}\left [1+W\left(-e^{-216J/ \rm{St}-1}\right)\right ],
\end{equation}
and a dimensionless lengthscale ratio by dividing a (``macroscopic'') lengthscale  $\mathcal{L}=d$ by $l_{\mu}$:
\begin{subequations}
\begin{equation}
    \mathcal{G}=\frac{\mathcal{L}}{l_{\mu}}=\frac{216J^{2}}{18J-I^{2}(1-e^{-18/\rm{St}})}=12(J+\lambda_{\rm{St}}I^2);
\end{equation}
\begin{equation}
\lambda_{\rm{St}}=\frac{(1-e^{-18/\rm{St}})}{18-\rm{St}\times (1-e^{-18/\rm{St}})}.
    \label{Eq:lambda}
    \end{equation}
\end{subequations}
We provide relationships between these new rheological parameters and previously proposed parameters, specifically, $I, J,$ and $K$ in Table \ref{tab:present}.
    \begin{table*}
    \caption{Near$-\phi_c$ dynamics and rheological parameters for viscous to inertial behaviors ($St=\rho_{s} d^2 \dot{\gamma}/\eta_f \approx 0 \to \infty$) }\label{tab:present}
    \begin{ruledtabular}
    \begin{tabular}{cccc}
    Relationships among: &$ K=J\times(1+\lambda_0 \rm{St})$ \cite{trulsson2012transition,Tapia_PhysRevLett.129.078001}& $G=12J\times$ & $\mathcal{G}=12J\times[1+\lambda(\rm{St}) \rm{St}$] \\
    $I, J, K, G, \mathcal{G}$ ($\rm{St}=I^2/J)$ &$\lambda_{0}=$0.635\cite{trulsson2012transition},0.1\cite{Tapia_PhysRevLett.129.078001},0.5\cite{amarsid2017viscoinertial} & $\left[1+\frac{\rm{St/J}}{216}\left\{1+W(-e^{-216/(\rm{St/J})-1})\right\}\right]$  & $\lambda(St)=\frac{(1-e^{-18/St})}{18-St(1-e^{-18/St})}$ \\
    \hline
    \hline
    Flow regimes & Published $\Delta \phi(P_p,\dot {\gamma})$ relationships & $G=t_{f}/\mathcal{T}$ & $\mathcal{G}=\mathcal{L}/l$\\
     \hline
     Inertial ($\rm{St}\to \infty$)&$(\frac{\phi_{c}}{\phi}-1) \propto I\propto \frac{\dot{\gamma}}{\sqrt{P_p}}$\ \cite{GDR2004On,da2005rheophysics,jop2006constitutive}    & $(\frac{\phi_{c}}{\phi}-1) \propto \lim\limits_{\rm{St}/J \to \infty}G=\frac{2}{\sqrt3} I$ & $(\frac{\phi_{c}}{\phi}-1) \propto\lim\limits_{St \to \infty}\sqrt{\mathcal{G}}=\frac{2}{\sqrt 3}I$\\
    \hline
    Viscous ($\rm{St}\to 0$) &$(\frac{\phi_{c}}{\phi}-1) \propto \sqrt{ J} \propto \sqrt{\frac{\dot{\gamma}}{P_p}}$\ \cite{cassar2005submarine,boyer2011unifying}     & $(\frac{\phi_{c}}{\phi}-1)\propto \lim\limits_{\rm{St}/J \to 0}\sqrt G=\sqrt{12J}$ & $(\frac{\phi_{c}}{\phi}-1) \propto \lim\limits_{St \to 0}\sqrt{\mathcal{G}}=\sqrt{12J}$\\
    \hline
    Visco-inertial &$\phi_{c}-\phi \propto \sqrt K\ $ \cite{trulsson2012transition,Tapia_PhysRevLett.129.078001}\  & $(\frac{\phi_{c}}{\phi}-1) \propto G^{\alpha}$  & $(\frac{\phi_{c}}{\phi}-1) \propto \mathcal{G}^{0.5}$ \\
    ($\rm{St}\approx 0 \to \infty$) & $ (\frac{\phi_{c}}{\phi}-1) \propto \sqrt K\ $ \cite{amarsid2017viscoinertial} & $ \alpha=0.5\to 1$ &  \\
    \end{tabular}
    \end{ruledtabular}
    \end{table*} 

\begin{figure*}
    	\centering
    	\includegraphics[scale=0.41]{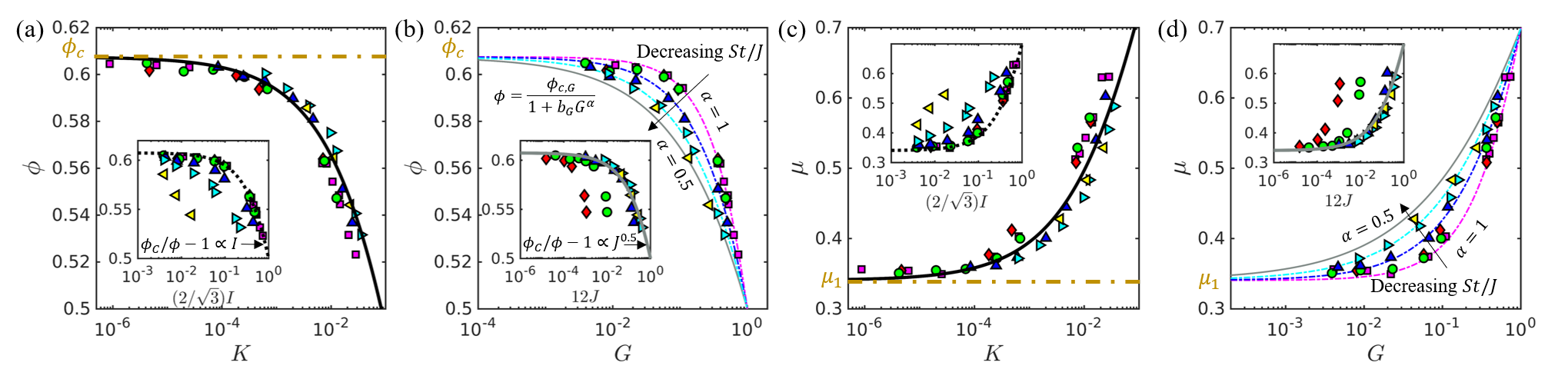}
    	\caption{DEM data for $\phi$ and $\mu$ plotted vs. $K$ and $G$ for particles submerged in fluids of different viscosities [$\eta_{f}=$0.1 cP ($\Diamond$), 1 cP ($\bigcirc$), 20 cP ($\Delta$), 100 cP ($\rhd$), 10$^{3}$ cP ($\lhd$), and 0 ($\Box$)]. (a) $\phi$ vs. $K=J+\lambda_o I^2$, with $\lambda_o=0.1$, (inset $\phi$ vs. $2I/\sqrt{3}$, which is ${t_{\mu}}/{\mathcal{T}}$ in dry conditions). (b) $\phi$ vs.\ ${t_{\mu}}/{\mathcal{T}}=G$ (inset $\phi$ vs.\ $12J$,  ${t_{\mu}}/{\mathcal{T}}$, $G$ in extremely viscous condition.). (c) $\mu$ vs.\ $K$ (inset $\mu$ vs.\  $(2/\sqrt{3})I$). (d) $\mu$ vs.\ $G$  (inset $\mu$ vs.\ $12J$), For (b) and (d), We note the systematic change in the functional forms of $(\phi_{c}/\phi-1)$ vs.\ $G$, and $(\mu-\mu_{c})$ vs. $G$, as St/J decreases from high to low and $G$ transforms from $(2/\sqrt{3})I$ to $12J$. The lines are fitted using Eqn.\ \ref{Eq:MU}-\ref{Eq:PHI}; the fit parameters for $\mathcal{X}=K$ in (a) and (c) are in Table 2.  For $\mathcal{X}=G$ in (b) and (d),$\mathcal{X}=2I/\sqrt{3}$ inset of (a),(c), and $\mathcal{X}=12J$ inset of (b), (d) the fit parameters are $\phi_{c,\mathcal{X}}$=0.6075, $b_{\mathcal{X}}$=0.215, $\mu_{1,\mathcal{X}}=$0.34, $\mu_{2,\mathcal{X}}=$1.49, and $Q_{\mathcal{X}}=$2.2. $\alpha$ for $G$ decreases from 1 to 0.5 as St/J decreases from high to low, $\alpha=1$ for $2I/\sqrt{3}$ and $\alpha=0.5$ for $12J$.}
    	\label{fig:fitsGK}
\end{figure*}

To test the relative effectiveness of $K$, $\mathcal{G}$, and $G$ in capturing the visco-inertial behaviors of granular fluid flows, we use new 3D Discrete Element Method (DEM) simulation data.  Our DEM particles are governed by previously published relationships for (1) linear interparticle contact forces \cite{cundall1979discrete}, (2) a viscous (Stokes) fluid drag\cite{trulsson2012transition}, and (3) lubrication forces\cite{goldman1967slow, man2018rheology}, for completeness, in the supplement.  Each simulation uses 2800 spheres whose radii are uniformly distributed between 0.07$-$0.1 cm.  Particle density $\rho_{s}$= 1.18 g/cm$^3$; normal and tangential stiffnesses are $5\times10^{7}$ dyn/cm, and $2.5\times10^{7}$ dyn/cm, respectively, and their restitution and frictional coefficients are 0.2 and 0.3, respectively. Our particles are bounded by two rough plates normal to the $z-$ direction and periodic boundaries in both $x-$ and $y-$ directions  (Fig.\ref{setup}(a)).  The two plates move with equal and opposing $x-$ velocities. The bottom plate is fixed in the $z-$direction, while the $z-$position of the top plate adjusts to exert constant normal stress ($P_p$) to the particles.  From one simulation to the next, we vary the dynamics via the shear rate $\dot \gamma=$ 0.15 to 50 s$^{-1}$, the confining pressure $P_p=$ 80 to 300 Pa, and the fluid viscosity $\eta _f=$ 0.1 to 1000 cP.   We calculate the interparticle stress tensor $\sigma_{ij}$ through all contact pairs \cite{goldhirsch2002microscopic} and, from this, $P_p$, $\tau$, and $\mu$.  We calculate $\phi$, by summing the particle volumes divided by the cell volume. For both $\sigma_{ij}$ and $\phi$ we exclude regions adjacent to the plates to avoid associated inhomogeneities. 

\begin{figure}
	\centering
    	\includegraphics[scale=0.24]{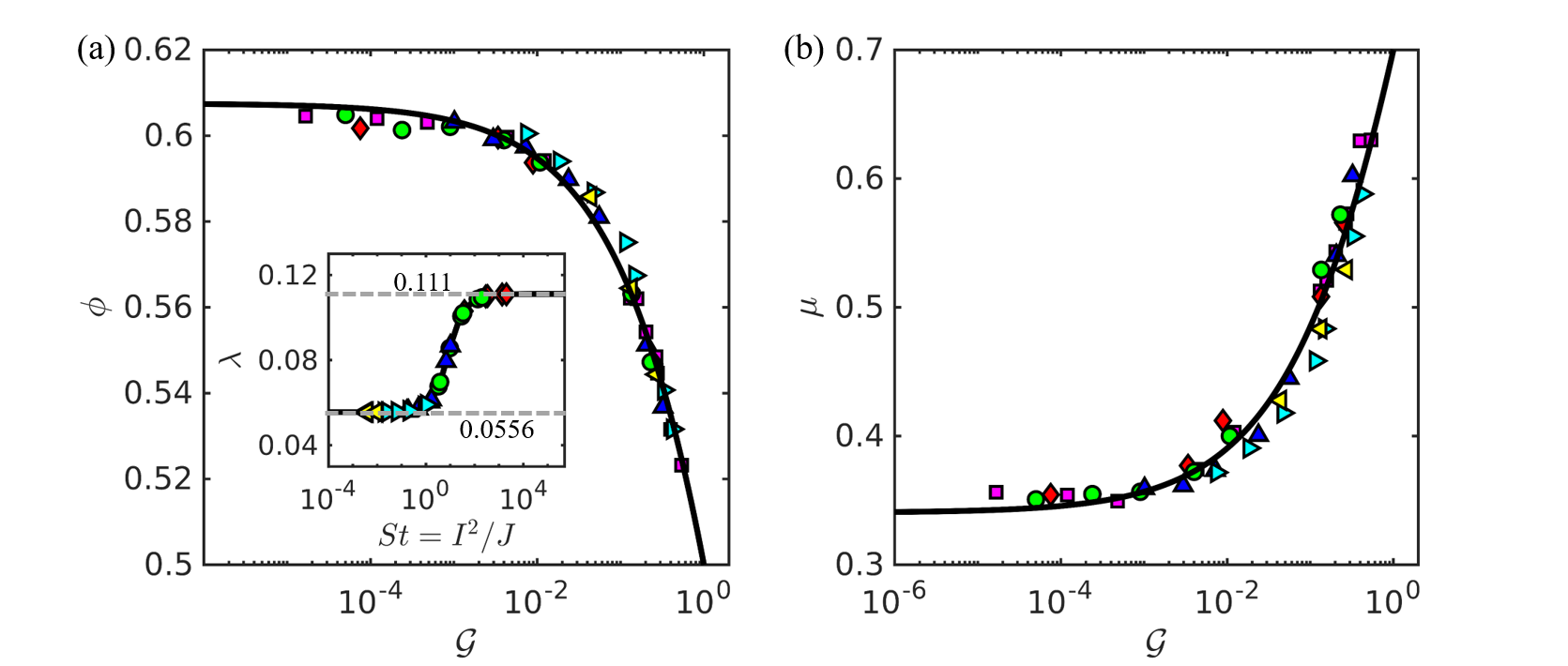}
    	\caption{We plot (a) $\phi$ and (b) $\mu$ as a function of the length scale $\mathcal{G}$. The solid lines in (a) and (b) are fitted by Eqs.\ref{Eq:MU} and \ref{Eq:MU}, respectively. Inset of (a) shows $\lambda$ vs.\ $St$ (Eqn.\ref{Eq:lambda}). }
	\label{fig:fitsmathcalG}
\end{figure}
We assess the rheologies once the system has reached a steady state, (Figs.\ref{setup}(b-d)) 
by plotting $\mu$ and $\phi$ vs.\ each of the three parameters $K, G,$ or $\mathcal{G}$ (Figs.\ \ref{fig:fitsGK}-\ref{fig:fitsmathcalG}).  In all three cases, the data can be fitted reasonably well by previously proposed relationships: 
\begin{subequations}
\begin{equation}
    \mu(\mathcal{X})=\mu_{1,\mathcal{X}}+\frac{\mu_{2,\mathcal{X}}-\mu_{1,\mathcal{X}}}{1+\mathcal{Q}_{\mathcal{X}}/\mathcal{X}^{\alpha}}, \rm{and} \label{Eq:MU}
\end{equation}
    \begin{equation}
    \phi(\mathcal{X})=\frac{\phi_{c,\mathcal{X}}}{1+b_\mathcal{X}\mathcal{X}^{\alpha}}.
    \label{Eq:PHI}
\end{equation}
\end{subequations}
Here,  $\phi_{c,\mathcal{X}}$, $b_{\mathcal{X}}$, $\mu_{1,\mathcal{X}}$, $\mu_{2,\mathcal{X}}$, and $\mathcal{Q}_{\mathcal{X}}$ are fit parameters for each rheological parameter (e.g., Table.\ref{tab:REG}) and each constant $\mathcal{Q}$ delineated with subscript $\mathcal{X}$, i.e.,  $``\mathcal{Q}_{\mathcal{X}}$'' refers to fit parameter $\mathcal{Q}$ obtained using $\mathcal{X}=K, G,$ or $\mathcal{G}$ in fitting Eqns. \ref{Eq:MU}-\ref{Eq:PHI} to the data. The form of $\mu$ indicates there is a lower and higher limit for $\mu$ associated with quasistatic and kinetic limits, respectively\cite{jop2006constitutive}.  The form of $\phi$ indicates only an upper limit for the dense flows, similar to a random close-packed configuration. 
Best fit coefficients (Tab \ref{tab:REG} and Fig.\ \ref{fig:fitsGK}) determined by regression analysis (detailed in Supplement) are similar for $K, G,$ and $\mathcal{G}$ (identical for $G$ and $\mathcal{G}$ derived consistently from Eqn.\ \ref{eq:equli}). 

In addition to the similarities among these rheological frameworks, we note two important differences that provide physical insights in the form of two variables in the expressions for $\mathcal{G}$ and $G$ ($\lambda$ and $\alpha$, respectively). Variation in $\alpha$ in the fits for $G$ demonstrate explicitly how the rheology changes from Newtonian to shear thickening as $St/J$ changes from $\approx \infty \to  0$ for the entire range of $\phi$ relevant for dense flows. In particular, the form recovers the earliest predictive rheological relationships for $\phi$ and $\mu$ vs. $I$ for dry particle flows, i.e., in the high $St/J$ limit of $G$ (Fig.\ \ref{fig:fitsGK} a,c ) and for $\phi$ and $\mu$ vs. $J$ for viscous-dominated flows, i.e., in the low $St/J$ limit of $G$ (Fig.\ \ref{fig:fitsGK} b,d ).  Variation in $\lambda(\rm{St})$ in the fits for $\mathcal{G}$ provide an analogous transition.  At the same time, since $\mathcal{G}$ can be written in a similar form as $K$, with a theoretically predictable transition coefficient $\lambda(\rm{St})$, we can build on previous insights developed (e.g., Ref.\ \cite{trulsson2012transition,Tapia_PhysRevLett.129.078001} 
(Inset, Fig. 3(a)).   The singular value found to fit best with one experimental data set ($\lambda=0.1$) \cite{Tapia_PhysRevLett.129.078001} is contained in the range of $\lambda(\rm{St})$ found here.  Additionally, the predictions from $\mathcal{G}$ hold for other systems with other particle properties, when using appropriate material properties for each fit (e.g., in $\mu_1, \mu_2$, and $\phi_c$, as argued in Ref.\ \cite{man2023friction}).  Examples include 3-d data \cite{Tapia_PhysRevLett.129.078001,boyer2011unifying,savage1984289,WOS:000331956500014} (Fig.\ref{stlambda}) and 2-d systems \cite{amarsid2017viscoinertial} the latter found in the Supplementary file.  The fits of previously published data (Fig.\ 4 and Supplementary) indicate that the coefficients $\mathcal{Q}_{\mathcal{G}}$ in $\mu$ and $b_{\mathcal{G}}$ in $\phi$ are generally applicable for the rheology in all these systems while $\mu_1, \mu_2$, and $\phi_c$ are specific to each system (particle properties and boundary conditions).  What determines the apparently universal values for $\mathcal{Q}_{\mathcal{G}}$ in $\mu$ and $b_{\mathcal{G}}$ in all of these systems is beyond the scope of this communication, though we are currently pursuing this issue. 

Before we conclude, we note that the question of appropriate system scales have important implications, so we briefly consider them  explicitly  here.  As we recall, $I$ and $J$ were derived using ratios of macro- to micro- timescales for inertial and viscous systems, respectively, while the intentionally-designed cross-rheology parameter $K$ was initially proposed based on a linear superposition of stress scales.\cite{trulsson2012transition} (as was a similarly proposed form $I_m=\sqrt{\alpha_1 I^2+\alpha_2 J}$\cite{amarsid2017viscoinertial}).  The success of the forms based on stress scales and the commonality of the pressure scaling of near-$\phi_c$ dependence (e.g., Tab 1) implicates stresses, rather than time scales, as intrinsic to the scaling of these systems.   Still, starting from these considerations without first-principles guidance leads to an additional fit parameter ($\lambda_o$ for K or $\alpha_1 / \alpha_2$ for $I_m$) whose value varies from one system to the next.  Our new timescale parameter that treats the systems as ratios between timescales from first principles introduces a different variable, $\alpha$ which varies from 1/2 to 1 as the system changes from viscous to inertial (Tab \ref{tab:present}).  This is not surprising as, under constant pressure boundary conditions, it forces a rheology dependent on $\dot{\gamma}$ in the viscous limit to transition to one dependent on $\dot{\gamma}^2$ in the inertial limit, shown previously to hold for these systems\cite{boyer2011unifying,Tapia_PhysRevLett.129.078001}.  Our lengthscale-based parameter $\mathcal{G}$ eliminates any such case-by-case fit parameter.
\begin{table}[b]
\caption{\label{tab:REG}%
Regression analysis for the expression of $\mu$ and $\phi$.}
\begin{ruledtabular}
\begin{tabular}{cccc}
Expression&Fitting parameter&R-Square&
RMSE\\
\hline
$\phi_{c}/(1+b\sqrt{\mathcal{G}})$&$b$, $\phi_{c}$=0.215,0.6075&0.99&3.0e-3\\
$\phi_{c}/(1+b\sqrt{K})$ & $b$, $\phi_{c}$=0.826,0.6072 & 0.95 & 5.8e-3 \\
$\mu_{1}+\frac{\mu_{2}-\mu_{1}}{1+\mathcal{Q}/\sqrt{\mathcal{G}}}$
  &$\mathcal{Q}$,$\mu_{1}$,$\mu_{2}$=2.2, 0.34, 1.49
  & 0.97 & 1.6e-2\\
$\mu_{1}+\frac{\mu_{2}-\mu_{1}}{1+\mathcal{Q}/\sqrt{K}}$
  &$\mathcal{Q}$,$\mu_{1}$,$\mu_{2}$=0.45, 0.34, 1.29& 0.91 & 2.8e-2\\
\end{tabular}
\end{ruledtabular}
\end{table}
To add to these mathematically-based considerations, we consider related evidence suggesting the importance of lengthscales and the physics-based behaviors of these dense particulate mixtures.   In addition to the rheological studies reviewed here, structural studies of near-jammed (near $\phi_c$) systems have shown the importance of both small- and large- lengthscale structures play an important role for particle transport (e.g., Refs.\  \cite{choi2004diffusion,bocquet2009kinetic, bonnoit2010mesoscopic}.  For example, Choi et al. (2004) \cite{choi2004diffusion} showed that near-$\phi_c$ systems relied on dynamics of cage-breaking, largely related to local ``holes'' into which otherwise caged particles could diffuse.  Additionally, the dynamical processes of the granular flows are strongly correlated to larger-scale correlated movements within a granular assembly.  While the relationships between the structures and flows remain open questions, considering the combined importance of theoretical lengthscales associated with structure and particle displament seems to hold important clues.

\begin{figure}
	\centering
	\includegraphics[scale=0.215]{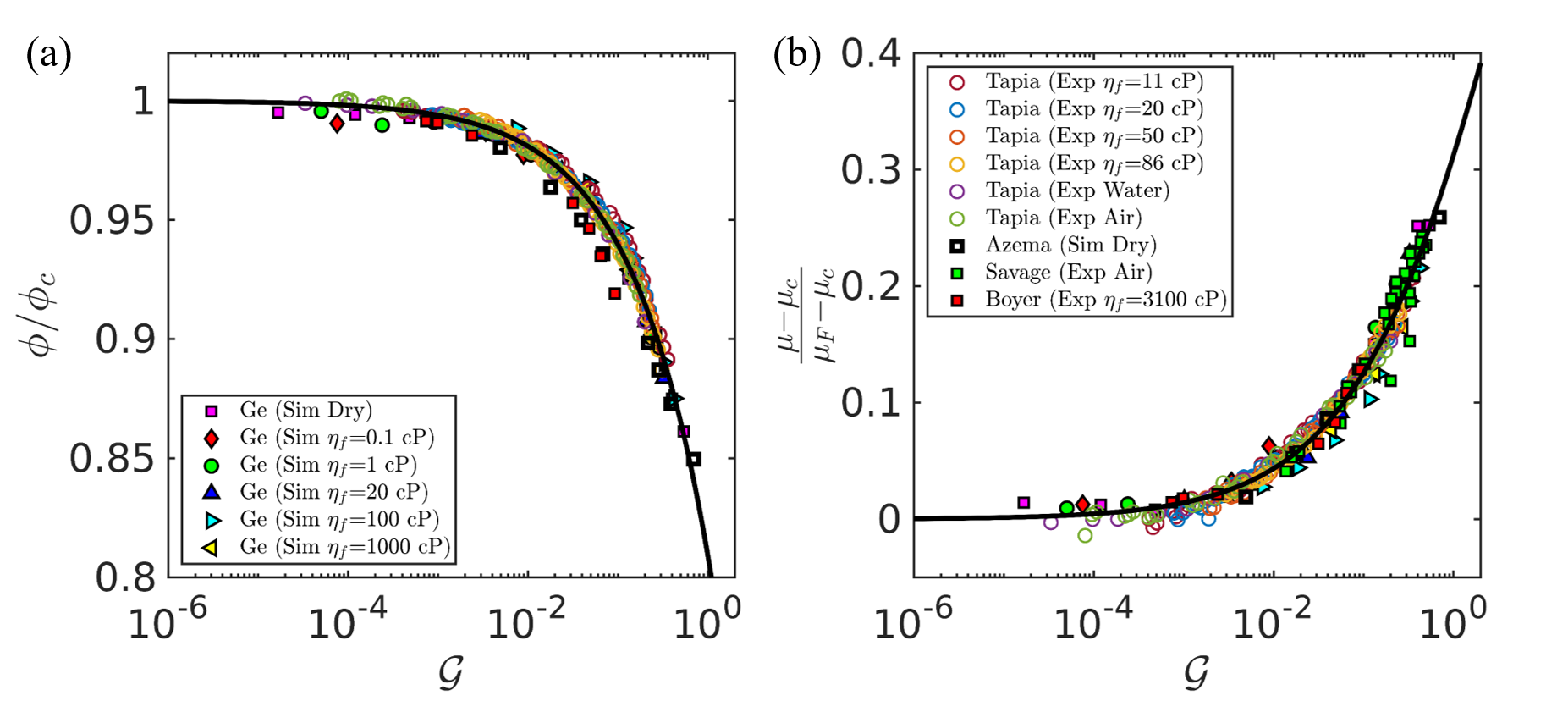}
	\caption{ Normalized solid fraction (a) and apparent frictional coefficient (b) plotted vs.\ data from $\mathcal{G}$ for data from Refs.\cite{Tapia_PhysRevLett.129.078001,boyer2011unifying,savage1984289,WOS:000331956500014}. To account for the varied material properties used in each study (reflected in $\mu_{1}, \mu_{2}$, and $\phi_{c}$) we plot $\phi/\phi_{c}$, and $\frac{\mu-\mu_{1}}{\mu_{2}-\mu_{1}}$ vs.\ $\mathcal{G}$. The solid line in (a) and (b) are from  Eqns.\ref{Eq:MU} and \ref{Eq:PHI},  using $b_{\mathcal{G}}$ and $\mathcal
 {Q}_{\mathcal{G}}$ from our data (Table 2).}
	\label{stlambda}
\end{figure}

To summarize, in this communication, we considered the inertial response of a sphere subjected to contact pressure $P_p$ and a linear drag force, and derived a particle-response micro-timescale $t_{\mu}=t(l=d)$ and equivalent micro-lengthscale $l_{\mu}=l(t=1/\dot{\gamma})$  (i.e., the time for a particle to travel by $d$ and the distance for a particle to travel in time $1/\dot{\gamma}$, respectively,  in response to this forcing). From these, we derived a timescale ratio and a lengthscale ratio: $G=t_{\mu}/ (1/ \dot{\gamma})$ and $\mathcal{G}=d/l_{\mu}$. When considering these in the context of new and previously published data, we found that the lengthscale ratio (rather than a timescale ratio) is intrinsically related to the rheology and requires no additional fit parameters.  
Based on these results along with recent work connecting particle properties with rheology parameters \cite{man2023friction}, we propose general constitutive relationships suitable for a wide range of dense, sheared granular flows across the viscous-inertial transition expressed using dimensionless scales $\mathcal{G}$ and $\rm{St}$  where the effects of viscous drag and inertia change under different confining pressures, fluid viscosities, and macroscopic deformations. These results provide insight toward expanding our understanding of the influence of different lengthscales in dense particle-fluid flows, providing greater promise for formulating constitutive models for larger-scale physics-based flow for larger scales than the ones we explored herein. 


This work is supported by the National Natural Science Foundation of China (NSFC grants NO. 12172305 and NO. 12202367) and the US National Science Foundation under grant number EAR-2127476. We thank Westlake High Performace Computing Center for computational resources and related assistance. The simulations were based on the MECHSYS open source library (\url{http://mechsys.nongnu.org}).

\nocite{*}

\bibliography{Hill_rheo}

\providecommand{\noopsort}[1]{}\providecommand{\singleletter}[1]{#1}%
\begin{thebibliography}{31}%
\makeatletter
\providecommand \@ifxundefined [1]{%
 \@ifx{#1\undefined}
}%
\providecommand \@ifnum [1]{%
 \ifnum #1\expandafter \@firstoftwo
 \else \expandafter \@secondoftwo
 \fi
}%
\providecommand \@ifx [1]{%
 \ifx #1\expandafter \@firstoftwo
 \else \expandafter \@secondoftwo
 \fi
}%
\providecommand \natexlab [1]{#1}%
\providecommand \enquote  [1]{``#1''}%
\providecommand \bibnamefont  [1]{#1}%
\providecommand \bibfnamefont [1]{#1}%
\providecommand \citenamefont [1]{#1}%
\providecommand \href@noop [0]{\@secondoftwo}%
\providecommand \href [0]{\begingroup \@sanitize@url \@href}%
\providecommand \@href[1]{\@@startlink{#1}\@@href}%
\providecommand \@@href[1]{\endgroup#1\@@endlink}%
\providecommand \@sanitize@url [0]{\catcode `\\12\catcode `\$12\catcode
  `\&12\catcode `\#12\catcode `\^12\catcode `\_12\catcode `\%12\relax}%
\providecommand \@@startlink[1]{}%
\providecommand \@@endlink[0]{}%
\providecommand \url  [0]{\begingroup\@sanitize@url \@url }%
\providecommand \@url [1]{\endgroup\@href {#1}{\urlprefix }}%
\providecommand \urlprefix  [0]{URL }%
\providecommand \Eprint [0]{\href }%
\providecommand \doibase [0]{https://doi.org/}%
\providecommand \selectlanguage [0]{\@gobble}%
\providecommand \bibinfo  [0]{\@secondoftwo}%
\providecommand \bibfield  [0]{\@secondoftwo}%
\providecommand \translation [1]{[#1]}%
\providecommand \BibitemOpen [0]{}%
\providecommand \bibitemStop [0]{}%
\providecommand \bibitemNoStop [0]{.\EOS\space}%
\providecommand \EOS [0]{\spacefactor3000\relax}%
\providecommand \BibitemShut  [1]{\csname bibitem#1\endcsname}%
\let\auto@bib@innerbib\@empty
\bibitem [{\citenamefont {Hutter}\ \emph {et~al.}(1995)\citenamefont {Hutter},
  \citenamefont {Koch}, \citenamefont {Pluüss},\ and\ \citenamefont
  {Savage}}]{1995The}%
  \BibitemOpen
  \bibfield  {author} {\bibinfo {author} {\bibfnamefont {K.}~\bibnamefont
  {Hutter}}, \bibinfo {author} {\bibfnamefont {T.}~\bibnamefont {Koch}},
  \bibinfo {author} {\bibfnamefont {C.}~\bibnamefont {Pluüss}},\ and\ \bibinfo
  {author} {\bibfnamefont {S.~B.}\ \bibnamefont {Savage}},\ }\bibfield  {title}
  {\bibinfo {title} {The dynamics of avalanches of granular materials from
  initiation to runout. part ii. experiments},\ }\href@noop {} {\bibfield
  {journal} {\bibinfo  {journal} {Acta Mechanica}\ }\textbf {\bibinfo {volume}
  {109}},\ \bibinfo {pages} {127} (\bibinfo {year} {1995})}\BibitemShut
  {NoStop}%
\bibitem [{\citenamefont {Tegzes}\ \emph {et~al.}(2002)\citenamefont {Tegzes},
  \citenamefont {Vicsek},\ and\ \citenamefont {Schiffer}}]{2002Avalanche}%
  \BibitemOpen
  \bibfield  {author} {\bibinfo {author} {\bibfnamefont {P.}~\bibnamefont
  {Tegzes}}, \bibinfo {author} {\bibfnamefont {T.}~\bibnamefont {Vicsek}},\
  and\ \bibinfo {author} {\bibfnamefont {P.}~\bibnamefont {Schiffer}},\
  }\bibfield  {title} {\bibinfo {title} {Avalanche dynamics in wet granular
  materials},\ }\href@noop {} {\bibfield  {journal} {\bibinfo  {journal}
  {Physical Review Letters}\ }\textbf {\bibinfo {volume} {89}},\ \bibinfo
  {pages} {094301} (\bibinfo {year} {2002})}\BibitemShut {NoStop}%
\bibitem [{\citenamefont {Yang}\ \emph {et~al.}(2020)\citenamefont {Yang},
  \citenamefont {Jing}, \citenamefont {Kwok},\ and\ \citenamefont
  {Sobral}}]{2020Pore}%
  \BibitemOpen
  \bibfield  {author} {\bibinfo {author} {\bibfnamefont {G.~C.}\ \bibnamefont
  {Yang}}, \bibinfo {author} {\bibfnamefont {L.}~\bibnamefont {Jing}}, \bibinfo
  {author} {\bibfnamefont {C.~Y.}\ \bibnamefont {Kwok}},\ and\ \bibinfo
  {author} {\bibfnamefont {Y.~D.}\ \bibnamefont {Sobral}},\ }\bibfield  {title}
  {\bibinfo {title} {Pore‐scale simulation of immersed granular collapse:
  Implications to submarine landslides},\ }\href@noop {} {\bibfield  {journal}
  {\bibinfo  {journal} {Journal of Geophysical Research: Earth Surface}\
  }\textbf {\bibinfo {volume} {125}} (\bibinfo {year} {2020})}\BibitemShut
  {NoStop}%
\bibitem [{\citenamefont {Kamrin}(2019)}]{kamrin2019non}%
  \BibitemOpen
  \bibfield  {author} {\bibinfo {author} {\bibfnamefont {K.}~\bibnamefont
  {Kamrin}},\ }\bibfield  {title} {\bibinfo {title} {Non-locality in granular
  flow: Phenomenology and modeling approaches},\ }\href@noop {} {\bibfield
  {journal} {\bibinfo  {journal} {Frontiers in Physics}\ }\textbf {\bibinfo
  {volume} {7}},\ \bibinfo {pages} {116} (\bibinfo {year} {2019})}\BibitemShut
  {NoStop}%
\bibitem [{\citenamefont {MiDi}(2004)}]{GDR2004On}%
  \BibitemOpen
  \bibfield  {author} {\bibinfo {author} {\bibfnamefont {G.}~\bibnamefont
  {MiDi}},\ }\bibfield  {title} {\bibinfo {title} {On dense granular flows},\
  }\href@noop {} {\bibfield  {journal} {\bibinfo  {journal} {The European
  Physical Journal E}\ }\textbf {\bibinfo {volume} {14}},\ \bibinfo {pages}
  {341} (\bibinfo {year} {2004})}\BibitemShut {NoStop}%
\bibitem [{\citenamefont {da~Cruz}\ \emph {et~al.}(2005)\citenamefont
  {da~Cruz}, \citenamefont {Emam}, \citenamefont {Prochnow}, \citenamefont
  {Roux},\ and\ \citenamefont {Chevoir}}]{da2005rheophysics}%
  \BibitemOpen
  \bibfield  {author} {\bibinfo {author} {\bibfnamefont {F.}~\bibnamefont
  {da~Cruz}}, \bibinfo {author} {\bibfnamefont {S.}~\bibnamefont {Emam}},
  \bibinfo {author} {\bibfnamefont {M.}~\bibnamefont {Prochnow}}, \bibinfo
  {author} {\bibfnamefont {J.-N.}\ \bibnamefont {Roux}},\ and\ \bibinfo
  {author} {\bibfnamefont {F.}~\bibnamefont {Chevoir}},\ }\bibfield  {title}
  {\bibinfo {title} {Rheophysics of dense granular materials: Discrete
  simulation of plane shear flows},\ }\href@noop {} {\bibfield  {journal}
  {\bibinfo  {journal} {Physical Review E}\ }\textbf {\bibinfo {volume} {72}},\
  \bibinfo {pages} {021309} (\bibinfo {year} {2005})}\BibitemShut {NoStop}%
\bibitem [{\citenamefont {Jop}\ \emph {et~al.}(2006)\citenamefont {Jop},
  \citenamefont {Forterre},\ and\ \citenamefont
  {Pouliquen}}]{jop2006constitutive}%
  \BibitemOpen
  \bibfield  {author} {\bibinfo {author} {\bibfnamefont {P.}~\bibnamefont
  {Jop}}, \bibinfo {author} {\bibfnamefont {Y.}~\bibnamefont {Forterre}},\ and\
  \bibinfo {author} {\bibfnamefont {O.}~\bibnamefont {Pouliquen}},\ }\bibfield
  {title} {\bibinfo {title} {A constitutive law for dense granular flows},\
  }\href@noop {} {\bibfield  {journal} {\bibinfo  {journal} {Nature}\ }\textbf
  {\bibinfo {volume} {441}},\ \bibinfo {pages} {727} (\bibinfo {year}
  {2006})}\BibitemShut {NoStop}%
\bibitem [{\citenamefont {Lacaze}\ and\ \citenamefont
  {Kerswell}(2009)}]{lacaze2009axisymmetric}%
  \BibitemOpen
  \bibfield  {author} {\bibinfo {author} {\bibfnamefont {L.}~\bibnamefont
  {Lacaze}}\ and\ \bibinfo {author} {\bibfnamefont {R.~R.}\ \bibnamefont
  {Kerswell}},\ }\bibfield  {title} {\bibinfo {title} {Axisymmetric granular
  collapse: a transient 3d flow test of viscoplasticity},\ }\href@noop {}
  {\bibfield  {journal} {\bibinfo  {journal} {Physical Review Letters}\
  }\textbf {\bibinfo {volume} {102}},\ \bibinfo {pages} {108305} (\bibinfo
  {year} {2009})}\BibitemShut {NoStop}%
\bibitem [{\citenamefont {Cassar}\ \emph {et~al.}(2005)\citenamefont {Cassar},
  \citenamefont {Nicolas},\ and\ \citenamefont
  {Pouliquen}}]{cassar2005submarine}%
  \BibitemOpen
  \bibfield  {author} {\bibinfo {author} {\bibfnamefont {C.}~\bibnamefont
  {Cassar}}, \bibinfo {author} {\bibfnamefont {M.}~\bibnamefont {Nicolas}},\
  and\ \bibinfo {author} {\bibfnamefont {O.}~\bibnamefont {Pouliquen}},\
  }\bibfield  {title} {\bibinfo {title} {Submarine granular flows down inclined
  planes},\ }\href@noop {} {\bibfield  {journal} {\bibinfo  {journal} {Physics
  of fluids}\ }\textbf {\bibinfo {volume} {17}},\ \bibinfo {pages} {103301}
  (\bibinfo {year} {2005})}\BibitemShut {NoStop}%
\bibitem [{\citenamefont {Boyer}\ \emph {et~al.}(2011)\citenamefont {Boyer},
  \citenamefont {Guazzelli},\ and\ \citenamefont
  {Pouliquen}}]{boyer2011unifying}%
  \BibitemOpen
  \bibfield  {author} {\bibinfo {author} {\bibfnamefont {F.}~\bibnamefont
  {Boyer}}, \bibinfo {author} {\bibfnamefont {{\'E}.}~\bibnamefont
  {Guazzelli}},\ and\ \bibinfo {author} {\bibfnamefont {O.}~\bibnamefont
  {Pouliquen}},\ }\bibfield  {title} {\bibinfo {title} {Unifying suspension and
  granular rheology},\ }\href@noop {} {\bibfield  {journal} {\bibinfo
  {journal} {Physical Review Letters}\ }\textbf {\bibinfo {volume} {107}},\
  \bibinfo {pages} {188301} (\bibinfo {year} {2011})}\BibitemShut {NoStop}%
\bibitem [{\citenamefont {Trulsson}\ \emph {et~al.}(2012)\citenamefont
  {Trulsson}, \citenamefont {Andreotti},\ and\ \citenamefont
  {Claudin}}]{trulsson2012transition}%
  \BibitemOpen
  \bibfield  {author} {\bibinfo {author} {\bibfnamefont {M.}~\bibnamefont
  {Trulsson}}, \bibinfo {author} {\bibfnamefont {B.}~\bibnamefont
  {Andreotti}},\ and\ \bibinfo {author} {\bibfnamefont {P.}~\bibnamefont
  {Claudin}},\ }\bibfield  {title} {\bibinfo {title} {Transition from the
  viscous to inertial regime in dense suspensions},\ }\href@noop {} {\bibfield
  {journal} {\bibinfo  {journal} {Physical Review Letters}\ }\textbf {\bibinfo
  {volume} {109}},\ \bibinfo {pages} {118305} (\bibinfo {year}
  {2012})}\BibitemShut {NoStop}%
\bibitem [{\citenamefont {Tapia}\ \emph {et~al.}(2022)\citenamefont {Tapia},
  \citenamefont {Ichihara}, \citenamefont {Pouliquen},\ and\ \citenamefont
  {Guazzelli}}]{Tapia_PhysRevLett.129.078001}%
  \BibitemOpen
  \bibfield  {author} {\bibinfo {author} {\bibfnamefont {F.}~\bibnamefont
  {Tapia}}, \bibinfo {author} {\bibfnamefont {M.}~\bibnamefont {Ichihara}},
  \bibinfo {author} {\bibfnamefont {O.}~\bibnamefont {Pouliquen}},\ and\
  \bibinfo {author} {\bibfnamefont {E.}~\bibnamefont {Guazzelli}},\ }\bibfield
  {title} {\bibinfo {title} {Viscous to inertial transition in dense granular
  suspension},\ }\href {https://doi.org/10.1103/PhysRevLett.129.078001}
  {\bibfield  {journal} {\bibinfo  {journal} {Physical Review Letters}\
  }\textbf {\bibinfo {volume} {129}},\ \bibinfo {pages} {078001} (\bibinfo
  {year} {2022})}\BibitemShut {NoStop}%
\bibitem [{\citenamefont {Man}\ \emph {et~al.}(2018)\citenamefont {Man},
  \citenamefont {Feng},\ and\ \citenamefont {Hill}}]{man2018rheology}%
  \BibitemOpen
  \bibfield  {author} {\bibinfo {author} {\bibfnamefont {T.}~\bibnamefont
  {Man}}, \bibinfo {author} {\bibfnamefont {Q.}~\bibnamefont {Feng}},\ and\
  \bibinfo {author} {\bibfnamefont {K.}~\bibnamefont {Hill}},\ }\bibfield
  {title} {\bibinfo {title} {Rheology of thickly-coated granular-fluid
  systems},\ }\href@noop {} {\bibfield  {journal} {\bibinfo  {journal} {arXiv
  preprint arXiv:1812.07083}\ } (\bibinfo {year} {2018})}\BibitemShut {NoStop}%
\bibitem [{\citenamefont {Bagnold}(1954)}]{bagnold1954experiments}%
  \BibitemOpen
  \bibfield  {author} {\bibinfo {author} {\bibfnamefont {R.~A.}\ \bibnamefont
  {Bagnold}},\ }\bibfield  {title} {\bibinfo {title} {Experiments on a
  gravity-free dispersion of large solid spheres in a newtonian fluid under
  shear},\ }\href@noop {} {\bibfield  {journal} {\bibinfo  {journal}
  {Proceedings of the Royal Society of London. Series A. Mathematical and
  Physical Sciences}\ }\textbf {\bibinfo {volume} {225}},\ \bibinfo {pages}
  {49} (\bibinfo {year} {1954})}\BibitemShut {NoStop}%
\bibitem [{\citenamefont {Bocquet}\ \emph {et~al.}(2009)\citenamefont
  {Bocquet}, \citenamefont {Colin},\ and\ \citenamefont
  {Ajdari}}]{bocquet2009kinetic}%
  \BibitemOpen
  \bibfield  {author} {\bibinfo {author} {\bibfnamefont {L.}~\bibnamefont
  {Bocquet}}, \bibinfo {author} {\bibfnamefont {A.}~\bibnamefont {Colin}},\
  and\ \bibinfo {author} {\bibfnamefont {A.}~\bibnamefont {Ajdari}},\
  }\bibfield  {title} {\bibinfo {title} {Kinetic theory of plastic flow in soft
  glassy materials},\ }\href@noop {} {\bibfield  {journal} {\bibinfo  {journal}
  {Physical review letters}\ }\textbf {\bibinfo {volume} {103}},\ \bibinfo
  {pages} {036001} (\bibinfo {year} {2009})}\BibitemShut {NoStop}%
\bibitem [{\citenamefont {Savage}(1984)}]{savage1984289}%
  \BibitemOpen
  \bibfield  {author} {\bibinfo {author} {\bibfnamefont {S.~B.}\ \bibnamefont
  {Savage}},\ }\bibfield  {title} {\bibinfo {title} {The mechanics of rapid
  granular flows},\ }\href@noop {} {\bibfield  {journal} {\bibinfo  {journal}
  {Advances in Applied Mechanics}\ }\textbf {\bibinfo {volume} {24}},\ \bibinfo
  {pages} {289} (\bibinfo {year} {1984})}\BibitemShut {NoStop}%
\bibitem [{\citenamefont {Az\'ema}\ and\ \citenamefont
  {Radja\"{\i}}(2014)}]{WOS:000331956500014}%
  \BibitemOpen
  \bibfield  {author} {\bibinfo {author} {\bibfnamefont {E.}~\bibnamefont
  {Az\'ema}}\ and\ \bibinfo {author} {\bibfnamefont {F.}~\bibnamefont
  {Radja\"{\i}}},\ }\bibfield  {title} {\bibinfo {title} {Internal structure of
  inertial granular flows},\ }\href
  {https://doi.org/10.1103/PhysRevLett.112.078001} {\bibfield  {journal}
  {\bibinfo  {journal} {Phys. Rev. Lett.}\ }\textbf {\bibinfo {volume} {112}},\
  \bibinfo {pages} {078001} (\bibinfo {year} {2014})}\BibitemShut {NoStop}%
\bibitem [{\citenamefont {Amarsid}\ \emph {et~al.}(2017)\citenamefont
  {Amarsid}, \citenamefont {Delenne}, \citenamefont {Mutabaruka}, \citenamefont
  {Monerie}, \citenamefont {Perales},\ and\ \citenamefont
  {Radjai}}]{amarsid2017viscoinertial}%
  \BibitemOpen
  \bibfield  {author} {\bibinfo {author} {\bibfnamefont {L.}~\bibnamefont
  {Amarsid}}, \bibinfo {author} {\bibfnamefont {J.-Y.}\ \bibnamefont
  {Delenne}}, \bibinfo {author} {\bibfnamefont {P.}~\bibnamefont {Mutabaruka}},
  \bibinfo {author} {\bibfnamefont {Y.}~\bibnamefont {Monerie}}, \bibinfo
  {author} {\bibfnamefont {F.}~\bibnamefont {Perales}},\ and\ \bibinfo {author}
  {\bibfnamefont {F.}~\bibnamefont {Radjai}},\ }\bibfield  {title} {\bibinfo
  {title} {Viscoinertial regime of immersed granular flows},\ }\href@noop {}
  {\bibfield  {journal} {\bibinfo  {journal} {Physical Review E}\ }\textbf
  {\bibinfo {volume} {96}},\ \bibinfo {pages} {012901} (\bibinfo {year}
  {2017})}\BibitemShut {NoStop}%
\bibitem [{\citenamefont {Cundall}\ and\ \citenamefont
  {Strack}(1979)}]{cundall1979discrete}%
  \BibitemOpen
  \bibfield  {author} {\bibinfo {author} {\bibfnamefont {P.~A.}\ \bibnamefont
  {Cundall}}\ and\ \bibinfo {author} {\bibfnamefont {O.~D.}\ \bibnamefont
  {Strack}},\ }\bibfield  {title} {\bibinfo {title} {A discrete numerical model
  for granular assemblies},\ }\href@noop {} {\bibfield  {journal} {\bibinfo
  {journal} {Geotechnique}\ }\textbf {\bibinfo {volume} {29}},\ \bibinfo
  {pages} {47} (\bibinfo {year} {1979})}\BibitemShut {NoStop}%
\bibitem [{\citenamefont {Goldman}\ \emph {et~al.}(1967)\citenamefont
  {Goldman}, \citenamefont {Cox},\ and\ \citenamefont
  {Brenner}}]{goldman1967slow}%
  \BibitemOpen
  \bibfield  {author} {\bibinfo {author} {\bibfnamefont {A.~J.}\ \bibnamefont
  {Goldman}}, \bibinfo {author} {\bibfnamefont {R.~G.}\ \bibnamefont {Cox}},\
  and\ \bibinfo {author} {\bibfnamefont {H.}~\bibnamefont {Brenner}},\
  }\bibfield  {title} {\bibinfo {title} {Slow viscous motion of a sphere
  parallel to a plane wall—i motion through a quiescent fluid},\ }\href@noop
  {} {\bibfield  {journal} {\bibinfo  {journal} {Chemical engineering science}\
  }\textbf {\bibinfo {volume} {22}},\ \bibinfo {pages} {637} (\bibinfo {year}
  {1967})}\BibitemShut {NoStop}%
\bibitem [{\citenamefont {Goldhirsch}\ and\ \citenamefont
  {Goldenberg}(2002)}]{goldhirsch2002microscopic}%
  \BibitemOpen
  \bibfield  {author} {\bibinfo {author} {\bibfnamefont {I.}~\bibnamefont
  {Goldhirsch}}\ and\ \bibinfo {author} {\bibfnamefont {C.}~\bibnamefont
  {Goldenberg}},\ }\bibfield  {title} {\bibinfo {title} {On the microscopic
  foundations of elasticity},\ }\href@noop {} {\bibfield  {journal} {\bibinfo
  {journal} {The European Physical Journal E}\ }\textbf {\bibinfo {volume}
  {9}},\ \bibinfo {pages} {245} (\bibinfo {year} {2002})}\BibitemShut {NoStop}%
\bibitem [{\citenamefont {Man}\ \emph {et~al.}(2023)\citenamefont {Man},
  \citenamefont {Zhang}, \citenamefont {Ge}, \citenamefont {Galindo-Torres},\
  and\ \citenamefont {Hill}}]{man2023friction}%
  \BibitemOpen
  \bibfield  {author} {\bibinfo {author} {\bibfnamefont {T.}~\bibnamefont
  {Man}}, \bibinfo {author} {\bibfnamefont {P.}~\bibnamefont {Zhang}}, \bibinfo
  {author} {\bibfnamefont {Z.}~\bibnamefont {Ge}}, \bibinfo {author}
  {\bibfnamefont {S.~A.}\ \bibnamefont {Galindo-Torres}},\ and\ \bibinfo
  {author} {\bibfnamefont {K.~M.}\ \bibnamefont {Hill}},\ }\bibfield  {title}
  {\bibinfo {title} {Friction-dependent rheology of dry granular systems},\
  }\href@noop {} {\bibfield  {journal} {\bibinfo  {journal} {Acta Mechanica
  Sinica}\ }\textbf {\bibinfo {volume} {39}},\ \bibinfo {pages} {1} (\bibinfo
  {year} {2023})}\BibitemShut {NoStop}%
\bibitem [{\citenamefont {Choi}\ \emph {et~al.}(2004)\citenamefont {Choi},
  \citenamefont {Kudrolli}, \citenamefont {Rosales},\ and\ \citenamefont
  {Bazant}}]{choi2004diffusion}%
  \BibitemOpen
  \bibfield  {author} {\bibinfo {author} {\bibfnamefont {J.}~\bibnamefont
  {Choi}}, \bibinfo {author} {\bibfnamefont {A.}~\bibnamefont {Kudrolli}},
  \bibinfo {author} {\bibfnamefont {R.~R.}\ \bibnamefont {Rosales}},\ and\
  \bibinfo {author} {\bibfnamefont {M.~Z.}\ \bibnamefont {Bazant}},\ }\bibfield
   {title} {\bibinfo {title} {Diffusion and mixing in gravity-driven dense
  granular flows},\ }\href@noop {} {\bibfield  {journal} {\bibinfo  {journal}
  {Physical review letters}\ }\textbf {\bibinfo {volume} {92}},\ \bibinfo
  {pages} {174301} (\bibinfo {year} {2004})}\BibitemShut {NoStop}%
\bibitem [{\citenamefont {Bonnoit}\ \emph {et~al.}(2010)\citenamefont
  {Bonnoit}, \citenamefont {Lanuza}, \citenamefont {Lindner},\ and\
  \citenamefont {Clement}}]{bonnoit2010mesoscopic}%
  \BibitemOpen
  \bibfield  {author} {\bibinfo {author} {\bibfnamefont {C.}~\bibnamefont
  {Bonnoit}}, \bibinfo {author} {\bibfnamefont {J.}~\bibnamefont {Lanuza}},
  \bibinfo {author} {\bibfnamefont {A.}~\bibnamefont {Lindner}},\ and\ \bibinfo
  {author} {\bibfnamefont {E.}~\bibnamefont {Clement}},\ }\bibfield  {title}
  {\bibinfo {title} {Mesoscopic length scale controls the rheology of dense
  suspensions},\ }\href@noop {} {\bibfield  {journal} {\bibinfo  {journal}
  {Physical review letters}\ }\textbf {\bibinfo {volume} {105}},\ \bibinfo
  {pages} {108302} (\bibinfo {year} {2010})}\BibitemShut {NoStop}%
\bibitem [{\citenamefont {Jaeger}\ \emph {et~al.}(1996)\citenamefont {Jaeger},
  \citenamefont {Nagel},\ and\ \citenamefont {Behringer}}]{jaeger1996granular}%
  \BibitemOpen
  \bibfield  {author} {\bibinfo {author} {\bibfnamefont {H.~M.}\ \bibnamefont
  {Jaeger}}, \bibinfo {author} {\bibfnamefont {S.~R.}\ \bibnamefont {Nagel}},\
  and\ \bibinfo {author} {\bibfnamefont {R.~P.}\ \bibnamefont {Behringer}},\
  }\bibfield  {title} {\bibinfo {title} {Granular solids, liquids, and gases},\
  }\href@noop {} {\bibfield  {journal} {\bibinfo  {journal} {Reviews of modern
  physics}\ }\textbf {\bibinfo {volume} {68}},\ \bibinfo {pages} {1259}
  (\bibinfo {year} {1996})}\BibitemShut {NoStop}%
\bibitem [{\citenamefont {Roux}\ and\ \citenamefont
  {Combe}(2002)}]{roux2002quasistatic}%
  \BibitemOpen
  \bibfield  {author} {\bibinfo {author} {\bibfnamefont {J.-N.}\ \bibnamefont
  {Roux}}\ and\ \bibinfo {author} {\bibfnamefont {G.}~\bibnamefont {Combe}},\
  }\bibfield  {title} {\bibinfo {title} {Quasistatic rheology and the origins
  of strain},\ }\href@noop {} {\bibfield  {journal} {\bibinfo  {journal}
  {Comptes Rendus Physique}\ }\textbf {\bibinfo {volume} {3}},\ \bibinfo
  {pages} {131} (\bibinfo {year} {2002})}\BibitemShut {NoStop}%
\bibitem [{\citenamefont {Goldhirsch}(2003)}]{goldhirsch2003rapid}%
  \BibitemOpen
  \bibfield  {author} {\bibinfo {author} {\bibfnamefont {I.}~\bibnamefont
  {Goldhirsch}},\ }\bibfield  {title} {\bibinfo {title} {Rapid granular
  flows},\ }\href@noop {} {\bibfield  {journal} {\bibinfo  {journal} {Annual
  review of fluid mechanics}\ }\textbf {\bibinfo {volume} {35}},\ \bibinfo
  {pages} {267} (\bibinfo {year} {2003})}\BibitemShut {NoStop}%
\bibitem [{\citenamefont {Scott}\ \emph {et~al.}(2006)\citenamefont {Scott},
  \citenamefont {Mann},\ and\ \citenamefont {Martinez~Ii}}]{scott2006general}%
  \BibitemOpen
  \bibfield  {author} {\bibinfo {author} {\bibfnamefont {T.~C.}\ \bibnamefont
  {Scott}}, \bibinfo {author} {\bibfnamefont {R.}~\bibnamefont {Mann}},\ and\
  \bibinfo {author} {\bibfnamefont {R.~E.}\ \bibnamefont {Martinez~Ii}},\
  }\bibfield  {title} {\bibinfo {title} {General relativity and quantum
  mechanics: towards a generalization of the lambert w function a
  generalization of the lambert w function},\ }\href@noop {} {\bibfield
  {journal} {\bibinfo  {journal} {Applicable Algebra in Engineering,
  Communication and Computing}\ }\textbf {\bibinfo {volume} {17}},\ \bibinfo
  {pages} {41} (\bibinfo {year} {2006})}\BibitemShut {NoStop}%
\bibitem [{\citenamefont {Veberi{\v{c}}}(2012)}]{veberivc2012lambert}%
  \BibitemOpen
  \bibfield  {author} {\bibinfo {author} {\bibfnamefont {D.}~\bibnamefont
  {Veberi{\v{c}}}},\ }\bibfield  {title} {\bibinfo {title} {Lambert w function
  for applications in physics},\ }\href@noop {} {\bibfield  {journal} {\bibinfo
   {journal} {Computer Physics Communications}\ }\textbf {\bibinfo {volume}
  {183}},\ \bibinfo {pages} {2622} (\bibinfo {year} {2012})}\BibitemShut
  {NoStop}%
\bibitem [{\citenamefont {Ness}\ and\ \citenamefont
  {Sun}(2015)}]{ness2015flow}%
  \BibitemOpen
  \bibfield  {author} {\bibinfo {author} {\bibfnamefont {C.}~\bibnamefont
  {Ness}}\ and\ \bibinfo {author} {\bibfnamefont {J.}~\bibnamefont {Sun}},\
  }\bibfield  {title} {\bibinfo {title} {Flow regime transitions in dense
  non-brownian suspensions: Rheology, microstructural characterization, and
  constitutive modeling},\ }\href@noop {} {\bibfield  {journal} {\bibinfo
  {journal} {Physical Review E}\ }\textbf {\bibinfo {volume} {91}},\ \bibinfo
  {pages} {012201} (\bibinfo {year} {2015})}\BibitemShut {NoStop}%
\bibitem [{\citenamefont {DeGiuli}\ \emph {et~al.}(2015)\citenamefont
  {DeGiuli}, \citenamefont {D{\"u}ring}, \citenamefont {Lerner},\ and\
  \citenamefont {Wyart}}]{degiuli2015unified}%
  \BibitemOpen
  \bibfield  {author} {\bibinfo {author} {\bibfnamefont {E.}~\bibnamefont
  {DeGiuli}}, \bibinfo {author} {\bibfnamefont {G.}~\bibnamefont {D{\"u}ring}},
  \bibinfo {author} {\bibfnamefont {E.}~\bibnamefont {Lerner}},\ and\ \bibinfo
  {author} {\bibfnamefont {M.}~\bibnamefont {Wyart}},\ }\bibfield  {title}
  {\bibinfo {title} {Unified theory of inertial granular flows and non-brownian
  suspensions},\ }\href@noop {} {\bibfield  {journal} {\bibinfo  {journal}
  {Physical Review E}\ }\textbf {\bibinfo {volume} {91}},\ \bibinfo {pages}
  {062206} (\bibinfo {year} {2015})}\BibitemShut {NoStop}%
\end{thebibliography}%


\providecommand{\noopsort}[1]{}\providecommand{\singleletter}[1]{#1}%
%
\end{document}